# High-temperature structural phase transition and infrared dielectric features of La$_2$CoMnO$_6$

R. X. Silva[1], A. Nonato[2], R. L. Moreira[3], R. M. Almeida[4] and C. W. A. Paschoal[5,*]

[1] Coordenação de Ciências Naturais, Universidade Federal do Maranhão, Campus VII, 65400-000, Codó - MA, Brazil

[2] Coordenação de Ciências Naturais, Universidade Federal do Maranhão, Campus do Bacabal, 65700-000, Bacabal - MA, Brazil

[3] Departamento de Física, Universidade Federal de Minas Gerais, ICEx, 31270-901 Belo Horizonte - MG, Brazil

[4] Departamento de Ensino Superior, Instituto Federal de Educação Ciência e Tecnologia do Maranhão, Campus São Luís - Maracanã, 65095-460, São Luís - MA, Brazil

[5] Departamento de Física, Universidade Federal do Ceará, Campus do Pici, 65455-900, Fortaleza - CE, Brazil

* Corresponding author. Tel: +55 85 3366 9908

E-mail address: paschoal.william@fisica.ufc.br (C. W. A Paschoal)

## Abstract

Temperature-dependent FAR-infrared reflectivity spectra of partially ordered magnetodielectric La$_2$CoMnO$_6$ is presented, from room temperature up to 675 K. A clear first-ordered structural phase transition (SPT) from a monoclinic structure with $P2_1/n$ symmetry to a rhombohedral phase with $R\bar{3}$ symmetry was evidenced from the behaviour of polar phonon modes at T$_C$ ~ 590 K. The temperature dependences of the transversal and longitudinal phonon branches, dielectric strengths, and damping of the strongest dielectric modes confirm the significant contribution of the phonon modes on the SPT, and revealed an important lattice anharmonicity, particularly for the low frequency modes. In addition, these investigations showed that structural ordering does not inhibit the SPT, and provided valuable information towards the polar phonons, their implications on intrinsic dielectric constant in double perovskites and in related compounds.

# 1. Introduction

Multiferroic materials, which have been studied since the middle of the last century, have attracted much attention in the last 15 years due to their physical properties and the enormous potential for technological applications, which involve sensors, microwave devices, thermal energy capture, photovoltaic cells, solid state cooling, data recording and storage, and random access memories, among many others[1–5]. Among multiferroics, double perovskites with the general formula RE$_2$MeMnO$_6$ (RE = rare earth elements and Me = Ni or Co) have attracted considerable attention since Rogado *et al.*[6] observed a near-room temperature magnetodielectric effect on La$_2$NiMnO$_6$. RE$_2$MeMnO$_6$ exhibit a variety of interesting and useful electronic and magnetic properties. Usually, at room temperature, they are characterized by a noncentrosymmetric monoclinic structure belonging to $P2_1/n$ (ITA number #14 or $C_{2h}^5$) space group, where the charge, the size and the ordering of the B-site have strong influence on their physical properties, as multiferroic behavior[7,8], magnetodielectric effect[9,10], magnetoresistance[11,12], spin-glass[13,14], among others. Besides, lattice parameter variations by means of the lanthanide contraction modify the degree of octahedral distortions, and consequently, change the B-O-B' bond angles, wherein are established the superexchange interactions which dominate the magnetism of these materials.

Bull *et al.*[15] were the first to report a charge-ordered monoclinic structure in La$_2$CoMnO$_6$ (LCMO) by using powder neutron diffraction, wherein the Co and Mn cations are arranged in a *rock salt* type structure, occupying the 2c and 2d Wyckoff sites, respectively. In addition, at high temperatures LCMO undergoes a structural phase transition (SPT) from the monoclinic phase to a rhombohedral $R\bar{3}$ (ITA number #148 or $S_6^2$) symmetry, at around 598 K, maintaining the charge order during this transformation. Iliev *et al.*[16] carried out Raman spectra measurements on epitaxial thin film of LCMO at high temperatures, but they did not observe the mentioned $P2_1/n \to R\bar{3}$ transition, when heating the sample up to 800 K. Recently, Kumar and Shate[17] studied the Raman active modes in bulk compound and epitaxial thin films, in the range between 300 K and 823 K. The appearance of new modes at 523 K and the anomalous changes in the stretching phonon parameters at around 583 K confirmed the structural transformation. Also, they considered that the structural transition in the LCMO was only detected because the samples were well ordered. Orayech *et al.*[18] reported a detailed study of the SPT sequence of the LCMO by using high-temperature X-ray powder diffraction in the temperature range between 300 and 1600 K, evidencing the effective distortion of the crystal cell from the highest symmetry. The first-order phase-transition from a monoclinic to a trigonal symmetry ($P2_1/n \to R\bar{3}$) was observed at 450 K, and the second phase transition, from the intermediate rhombohedral symmetry to a cubic phase ($Fm\bar{3}m$ or $O_h^5$ – ITA number #225), was observed at about 1545 K.

Raman spectroscopy is a non-destructive technique that has been extensively applied to study various double perovskites to detect spin-phonon coupling, SPT, crystallinity, composition, lattice distortions, and even magnetic inhomogeneities[16,17,19–24]. However, infrared (IR) spectroscopy, apart from the phononic response, allows the determination of the intrinsic dielectric properties of the materials and their temperature dependence, such as the intrinsic static dielectric constant ($\varepsilon_s$), the contribution of each IR-active phonon to the dielectric constant, the intrinsic dielectric losses (*tan δ*) and the unloaded quality factor ($Q_u$)[25–31]. All these parameters bring relevant information to investigate the microscopic mechanisms involved in structural phase transitions. Recently, we have investigated the intrinsic dielectric constant in LCMO by means of their IR-active optical phonons, which shed light on the discussion about the colossal dielectric constant in this material. In short, it was undoubtedly showed that the CDC has an extrinsic origin[32], since the intrinsic dielectric constant remains well behaved ($\varepsilon_s$ < 15), even for fully ordered samples[29,32]. Moreover, since all the crystal structures presented by LCMO are centrosymmetric, the IR and Raman-active optical phonons are complementary.

In this paper, we investigated the temperature dependence of the IR-active phonons of the partially ordered LCMO ceramic sample. From the temperature dependence of the phonon properties, we clearly observed the $P2_1/n \rightarrow R\bar{3}$ transition at around 590 K, showing that the structural order is not a s*ine qua non* condition to SPT in LCMO. The intrinsic dielectric properties are also discussed in the investigated temperature range.

## 2. Experimental procedures

La$_2$CoMnO$_6$ bulk samples were synthesized by modified polymeric precursor (MPP) route derived from the Pechini's method and calcined for 16h at 1100°C, as described in Ref [29]. X-ray diffraction data were collected by using X-ray powder diffraction (XRPD) on a Bruker D8 Advance equipped with a Cu-Ka radiation source (40 kV, 40 mA) over a range from 10° to 100° (0.02°/step with 0.3 s/step). The XRD pattern of the material calcined at 1100 ºC was indexed on a monoclinic unit cell with symmetry belonging to the space group $P2_1/n$, and lattice parameters, $a = 5.536$ Å, $b = 5.4875$ Å, $c = 7.7776$ Å and $\beta = 90.01°$. X-ray power diffraction pattern was compared with international diffraction database (ICSD# 98240)[15] using the GSAS code.[33]

Temperature dependence of the IR reflectivity spectra were acquired using a Fourier-transform spectrometer (Nicolet-Nexus 470) equipped with a Centaurus microscope (with a 10× objective and effective numerical aperture of 0.30), in the temperature range from 300 K to 673 K, using a Linkam TS1500 oven. In the mid-infrared region (550–4000 cm$^{-1}$), a SiC (Globar) lamp was used as infrared source, a Ge-coated KBr as beamsplitter and a LN$_2$-cooled HgCdTe was employed for the detection. In the far-infrared range (50–700 cm$^{-1}$), besides the same source, a Si-solid state beamsplitter and a LHe-cooled Si bolometer were used. In order to improve the reflectivity spectra, avoiding losses due to diffuse reflection at the rough sample surface, one region of the ceramic surface was covered with a thin gold coating that simulates a "rough" mirror used for the reference spectra. The measurements were performed under nitrogen purge, by accumulating 128 scans, with spectral resolution better than 2 cm$^{-1}$.

## 3. Results and discussions

Figure 1 shows the room temperature IR reflectivity spectrum of the partially ordered LCMO sample obtained at 1100 °C. The spectrum can be fitted according to the four-parameter semi-quantum model proposed by Gervais and Piriou[37] for describing the dielectric function. In this model, the complex dielectric function $\varepsilon(\omega)$ is expressed in terms of the IR-active phonons[31,32,34] as

$$\varepsilon(\omega) = \varepsilon_\infty \prod_j \frac{\left(\omega_{j,LO}^2 - \omega^2 + i\omega\gamma_{j,LO}\right)}{\left(\omega_{j,TO}^2 - \omega^2 + i\omega\gamma_{j,TO}\right)}, \quad (1)$$

where $\omega_{j,TO}$ and $\omega_{j,LO}$ correspond respectively to the frequencies of the transverse (TO) and longitudinal (LO) optical branches of the *j-th* vibrational mode, $\gamma_{j,TO}$ and $\gamma_{j,LO}$ are the corresponding damping constants of these branches, and $\varepsilon_\infty$ accounts for the contribution of the electronic polarization. At quasi-normal incidence, the dielectric function is related to the IR reflectivity by the Fresnel equation

$$R(\omega) = \left|\frac{\sqrt{\varepsilon(\omega)}-1}{\sqrt{\varepsilon(\omega)}+1}\right|^2. \qquad (2)$$

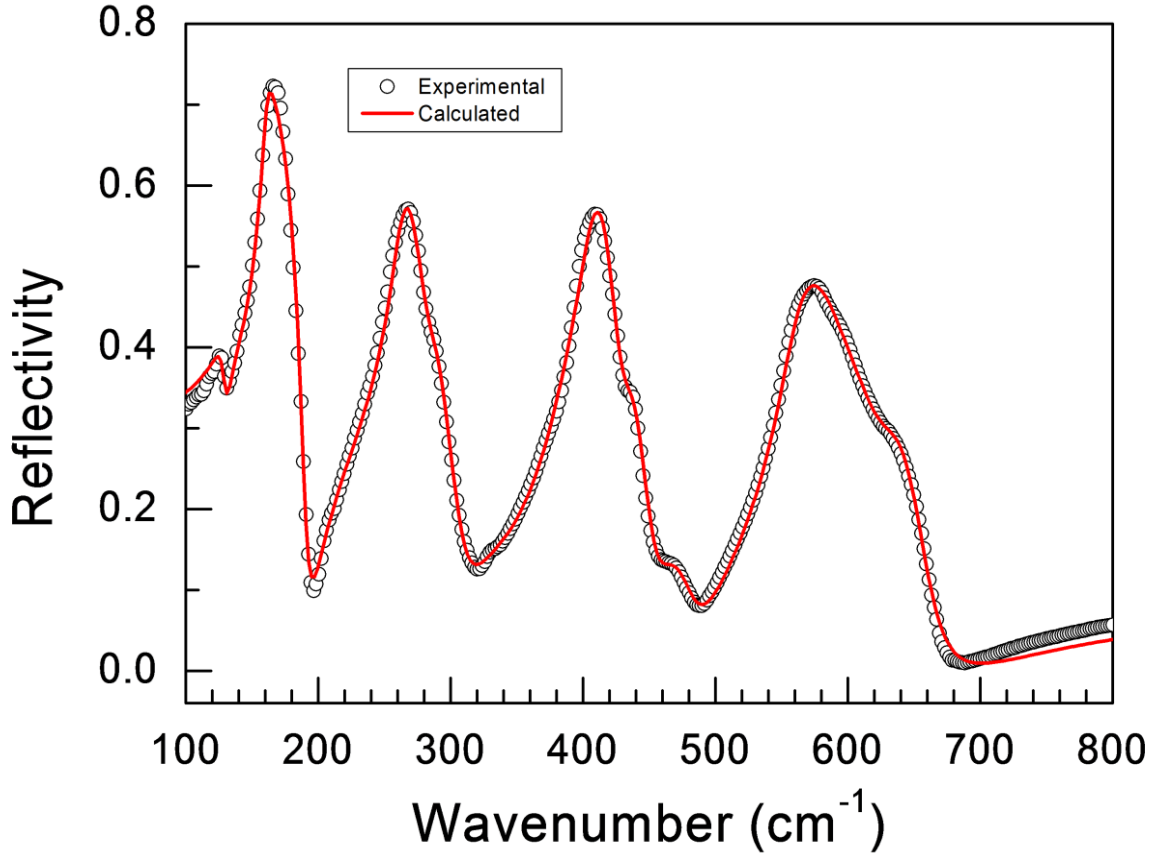

**Figure 1** – Infrared reflectivity spectrum of the partially ordered LCMO sample at room temperature. The black open circles represent the experimental data and red line represents the fitted spectrum by using the four-parameter semi-quantum model[34].

The red line in Figure 1 is the fitted curve for the reflectivity data of the LCMO (black open circles) in the spectral region covering the phonon modes, as obtained with the model outlined above. The adjustment parameters used are presented in Table I. This spectrum is quite similar to that one observed in a previous work for an ordered LCMO sample[29]. In both cases, the spectra are dominated by 4 broad bands that can be resolved (by fitting) into 13 polar modes.

**Table I** – Dispersion parameters from the best numerical fit of the FTIR spectrum of the LCMO ceramic at room temperature.

| Modes | $\omega_{j,TO}$ (cm$^{-1}$) | $\gamma_{j,TO}$ (cm$^{-1}$) | $\omega_{j,LO}$ (cm$^{-1}$) | $\gamma_{j,LO}$ (cm$^{-1}$) | $\Delta\varepsilon_{j,TO}$ | $10^8 \tan\delta_j/\omega$ |
|---|---|---|---|---|---|---|
| 1 | 128.7 | 9.4 | 129.9 | 7.7 | 0.35 | 1661 |
| 2 | 160.1 | 7.2 | 164.7 | 32.9 | 1.89 | 4521 |
| 3 | 170.5 | 37.3 | 190.9 | 12.8 | 1.53 | 16670 |
| 4 | 261.6 | 17.9 | 278.9 | 24.9 | 1.79 | 3970 |
| 5 | 284.5 | 30.7 | 306.0 | 31.4 | 0.29 | 949 |
| 6 | 342.3 | 58.5 | 344.8 | 53.3 | 0.080 | 336 |
| 7 | 388.6 | 37.6 | 389.7 | 46.7 | 0.10 | 214 |
| 8 | 402.6 | 24.7 | 425.9 | 23.5 | 0.91 | 1172 |
| 9 | 433.8 | 29.7 | 450.1 | 24.8 | 0.14 | 189 |
| 10 | 469.1 | 41.9 | 482.1 | 35.5 | 0.13 | 210 |
| 11 | 558.1 | 31.6 | 596.2 | 56.6 | 0.67 | 575 |
| 12 | 600.3 | 56.2 | 621.9 | 58.7 | 0.03 | 46 |
| 13 | 633.1 | 48.2 | 660.3 | 36.7 | 0.04 | 43 |
| $\varepsilon_\infty = 3.84$ | $\Sigma \Delta\varepsilon_j = 7.97$ | | | | $\Sigma \tan\delta_j/\omega = 18200 \times 10^{-8}$ | |
| $\varepsilon_s = \varepsilon_\infty + \Sigma \Delta\varepsilon_j = 11.81$ | | | | | $Q_u \times f = 164.5$ THz | |

As mentioned before, at room temperature LCMO crystallizes into a monoclinic structure belonging to the $P2_1/n$ (ITA number #14 or $C_{2h}^5$) space group. For this structure, the number and activities of the first-order vibrational modes can be predicted by using group theoretical tools, such as the nuclear-site method of Rousseau *et al.*[35], once the occupied Wyckoff positions are known[15,21]. Hence, the phonon modes at the $\Gamma$ point of the first Brillouin zone center in this phase can be distributed according to the $2/m$ ($C_{2h}$) factor group as shown in Table II.

**Table II** – Normal modes of vibration of the LCMO at the Brillouin zone center (Γ-point) in the monoclinic phase, with space group $P2_1/n$ (#14 or $C_{2h}^5$).

| Atom | Site | Site symmetry | Irreducible representations |
|---|---|---|---|
| La | 4e | $C_1$ | $3A_g \oplus 3A_u \oplus 3B_g \oplus 3B_u$ |
| Co | 2c | $C_i$ | $3A_u \oplus 3B_u$ |
| Mn | 2d | $C_i$ | $3A_u \oplus 3B_u$ |
| $O_{(1)}$ | 4e | $C_1$ | $3A_g \oplus 3A_u \oplus 3B_g \oplus 3B_u$ |
| $O_{(2)}$ | 4e | $C_1$ | $3A_g \oplus 3A_u \oplus 3B_g \oplus 3B_u$ |
| $O_{(3)}$ | 4e | $C_1$ | $3A_g \oplus 3A_u \oplus 3B_g \oplus 3B_u$ |

$$\Gamma_{IR} = 17A_u \oplus 16B_u$$
$$\Gamma_{Raman} = 12A_g \oplus 12B_g$$
$$\Gamma_{Acoustic} = A_u \oplus 2B_u$$

Therefore, according to the group theory predictions, 33 IR-active modes ($17A_u \oplus 16B_u$) are expected for monoclinic LCMO. The number of observed polar phonons is lower than that predicted by group theory, as it is usually observed in monoclinic ceramic samples[36]. Also, as discussed earlier[29], changes on cationic structural ordering mainly modify the phonon anharmonicity. The higher is the disorder of the samples, the wider are their bands at the spectrum, since the phonon lifetime along the lattice is low and the dielectric losses are large. This also pushes the intrinsic dielectric constant down[29,37]. However, the spectra of the ordered and partially disordered LCMO are very similar[29] and the modes can be assigned based on the work of Linka *et al.*[38]. Thus, we observe that the lowest frequency modes predominantly correspond to the (Co,Mn)–O beating, exhibiting a strong charge dynamical behavior and it is supposed to be sensitive to the B' and B'' sites content[32,36]. The second lowest frequency region (between 150 and 300 cm$^{-1}$) are associated with La–(Mn,Co)O$_6$ beating vibrations (A–BO$_6$ translation)[28,32], also named as Last-type modes. The intermediate region of the spectra are associated with the anti-phase vibration of the ions at the Co and Mn cation sites[28,38], whereas the highest wavenumber range, with $\omega > 500$ cm$^{-1}$, should be due to the octahedral stretching and bending modes (named the Axe-type modes)[25,28,32,39].

Also, from the phonon parameters, we can determine the dielectric mode strengths ($\Delta\varepsilon_j$), the high frequency permittivity ($\varepsilon_\infty$), the dielectric losses (*tan $\delta_j$*), as well as the extrapolated *static* dielectric constant ($\varepsilon_s$) and the quality factor ($Q_u$), at 10 GHz. The dielectric mode strengths are given by

$$\Delta\varepsilon_{j,TO} = \frac{\varepsilon_\infty}{\omega_{jTO}^2} \times \frac{\Pi_k\left(\omega_{k,LO}^2 - \omega_{j,TO}^2\right)}{\Pi_{k \neq j}\left(\omega_{k,TO}^2 - \omega_{j,LO}^2\right)}, \quad (3)$$

while the static dielectric constant has the following expression

$$\varepsilon_s = \varepsilon'(0) = \varepsilon_\infty + \sum \Delta\varepsilon_j, \quad (4)$$

and the dielectric losses can be calculated by

$$tan\,\delta_j = \omega \frac{\Delta\varepsilon_j \gamma_{j,TO}/\omega_{j,TO}^2}{\varepsilon_\infty + \sum_j \Delta\varepsilon_j}. \quad (5)$$

Finally, the intrinsic quality factor, $Q_u$, which is the reciprocal of the total phononic losses of the system, can be calculated by extrapolating to the microwave region (at 10 GHz, or 0.3333 cm$^{-1}$), and is presented as its product with frequency ($Q_u$ x *f*) in Table I, along with the aforementioned parameters. As expected, both ordered and partially ordered samples have similar microwave dielectric parameters[29,33]. The static dielectric constant is low, ca. 11.8, reinforcing that the CDC effect of the LCMO has extrinsic origin[32], even in partially ordered sample.

In addition, the calculated phonon parameters of Table I show that only five modes contribute more significantly to the intrinsic dielectric constant, the modes #2, #3, #4, #8, and #11, i.e., the highest dielectric strengths modes. Discarding the electronic polarization, the mode #2 ($\omega_{TO}$ = 160 cm$^{-1}$) has the strongest dielectric strength, with $\Delta\varepsilon_j$= 1.89, contributing approximately to 24% for the static (infrared) dielectric constant, against 46% of the same mode in an ordered sample[33]. In a reduced-ordered phase, this contribution was more diluted among the modes #2, #3 and #4, which together account for about 65% for the static dielectric constant, close to the 70% reported for the ordered sample. Therefore, the atomic vibrations associated with this band generate the highest dipole moment of the polar phonon spectrum of the LCMO and present the most significant changes during a phase transition, as it will be seen below. On the other hand, the contributions from other high dielectric strengths modes (modes #8 and #11) did not present significant changes, as compared to ordered samples[29,32].

Figure 2 shows the high-temperature far-IR reflectivity spectra of partially ordered LCMO. At high-temperature, LCMO can exhibit either rhombohedral or cubic structures. Table III summarizes the normal modes of vibration of the LCMO at the center of the Brillouin zone (Γ - point) and according to the group-theoretical analysis in an ideal cubic double ordered $Fm\bar{3}m$ ($O_h^5$ – ITA number #225) and $R\bar{3}$ ($S_6^2$ – ITA number #148) phases. Indeed, the four broad bands observed in the monoclinic phase spectrum are derived from the ideal ordered cubic phase, in which 4 distinct IR-active modes (4F$_{1u}$) are predicted if the octahedra tilts are absent. The symmetry reduction from the cubic phase and the increasing distortion promote the degeneracy splitting and the activation of additional first-order modes, enabling the visualization of new bands and shoulders for the systems with lower symmetries. Observe that, although the Wyckoff sites occupied by Mn and Co ions are different in the three phases, they have the same local symmetry in each structure ($Fm\bar{3}m$ → O$_h$, $R\bar{3}$ → S$_6$, P2$_1$/n → C$_i$), which do not contribute to the Raman spectra, but participate uniquely in the IR-active polar modes. Thus, for a deeper understanding of the contributions of these ions to lattice dynamics, information on the infrared-active modes are essential. Finally, it is worthy noticing that we expect only 10 IR-active phonons in the rhombohedral phase ($5A_u \oplus 5E_u$), according to Table III.

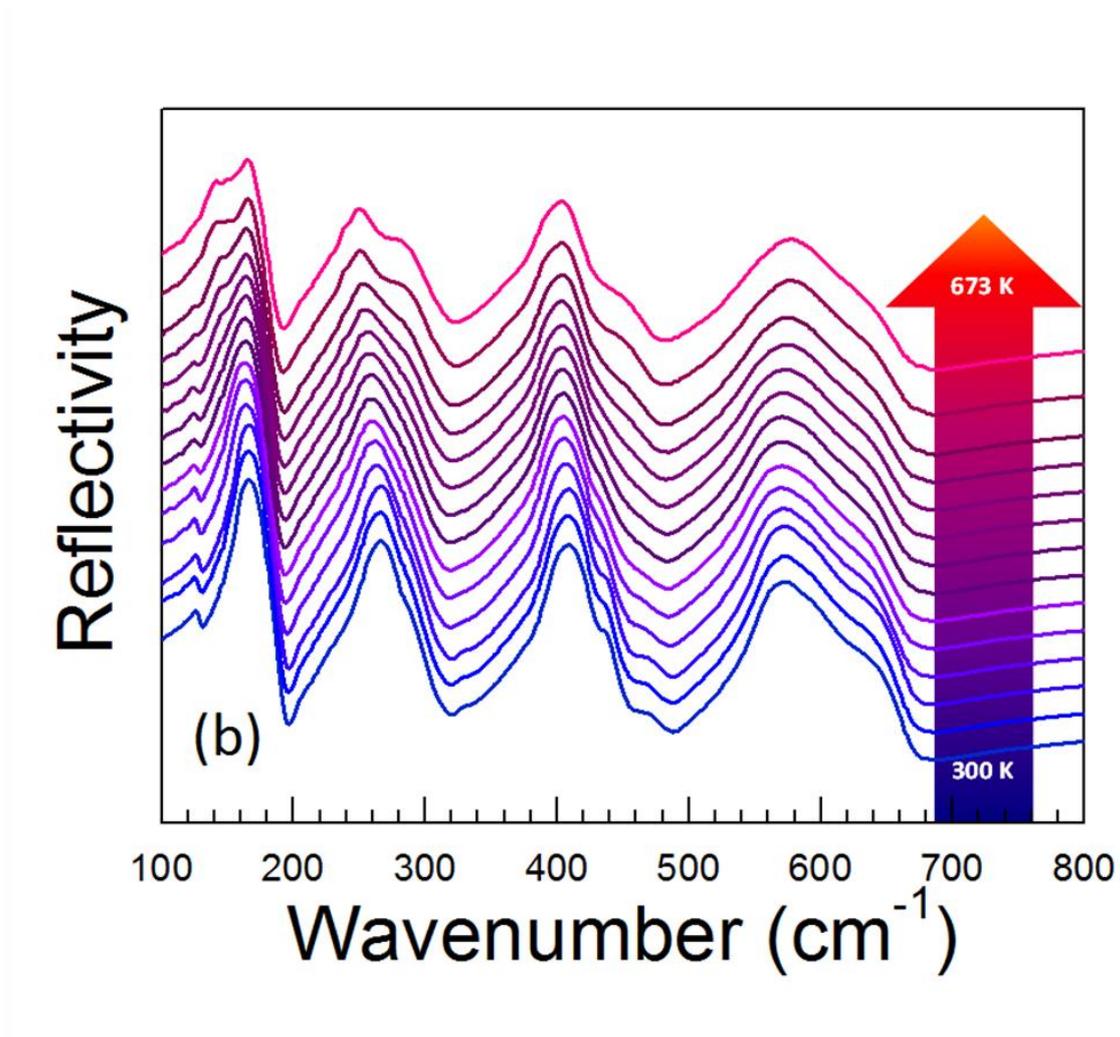

**Figure 2** – Temperature dependence of the far-IR reflectivity spectra of a partially ordered LCMO sample.

Figure 3 shows the temperature-dependence of the phonon positions (TO branches) observed in partially ordered LCMO. Clearly, the $P2_1/n \rightarrow R\bar{3}$ transition is observed at around 590 K, mainly marked by changes into lattice phonons at low-wavenumbers. The mode assignments allow us to draw qualitative conclusions concerning the influence of the phonon mode parameters on the SPT and on intrinsic dielectric properties. The critical temperature, $T_C$, observed is very close to the reported transition temperature for this transition (~ 598 K)[15]. The mode frequencies demonstrate kinks at $T_C$ followed by an anomalous softening or hardening on heating the ceramics. Moreover, the number of observed polar phonons reduces at temperatures above $T_C$. At high temperatures, the reflectivity spectra are better fitted with just 10 IR-active modes, as predicted by the group-theoretical analysis (5$A_u$ ⊕ 5$E_u$) for the expected rhombohedral phase.

**Table IIII** – Normal modes of vibration of the high-temperature LCMO phases at the Brillouin zone center (Γ-point).

| Structure | Atom | Site | Site symmetry | Irreducible representations |
|---|---|---|---|---|
| Cubic | La | 8c | $T_d$ | $F_{1u} \oplus F_{2g}$ |
| Fm-3m ($O_h^5$ - #225), Z=2 | Co | 4a | $O_h$ | $F_{1u}$ |
| | Mn | 4b | $O_h$ | $F_{1u}$ |
| | O | 28e | $C_{4v}$ | $A_{1g} \oplus E_g \oplus F_{1g} \oplus 2F_{1u} \oplus F_{2g} \oplus F_{2u}$ |
| | | | $\Gamma_{IR} = 4F_{1u}$ | |
| | | | $\Gamma_{Raman} = A_g \oplus E_g \oplus 2F_{2g}$ | |
| | | | $\Gamma_{Acoustic} = F_{1u}$ $\quad$ $\Gamma_{Silent} = F_{2u} \oplus F_{1g}$ | |
| Rhombohedral | La | 2c | $C_3$ | $A_g \oplus A_u \oplus E_g \oplus E_u$ |
| R-3 ($S_6^2$ - #148) | Co | 1a | $S_6$ | $A_u \oplus E_u$ |
| | Mn | 1b | $S_6$ | $A_u \oplus E_u$ |
| | O | 6f | $C_1$ | $3A_g \oplus 3A_u \oplus 3E_g \oplus 3E_u$ |
| | | | $\Gamma_{IR} = 5A_u \oplus 5E_u$ | |
| | | | $\Gamma_{Raman} = 4A_g \oplus 4E_g$ | |
| | | | $\Gamma_{Acoustic} = A_u \oplus E_u$ | |

Figure 4(a) displays the reflection spectra, at several temperatures, along with intensity maps for LCMO [Figure 4(b)], focused on the low-frequency spectral range between 120 cm$^{-1}$ and 230 cm$^{-1}$. As the temperature increases, the intensity of the modes decreases slowly and the modes naturally get broad due to thermal effects.

However, at the critical temperature the reflection intensity maps underwent significant changes, following a new trend in the distribution of intensities. The polar modes $\omega_{TO}\sim 160$ cm$^{-1}$ (mode #2) and $\omega_{TO}\sim 170$ cm$^{-1}$ (mode #3) demonstrate pronounced kinks, followed by strong softening behaviour of about 11 cm$^{-1}$ and 13 cm$^{-1}$ respectively, until the maximum temperature of 673 K [see Figure 4 (c)]. Notice that the damping constant of the mode #2 (at around 160 cm$^{-1}$) presents a peak in the critical region [Figure 4(d)].

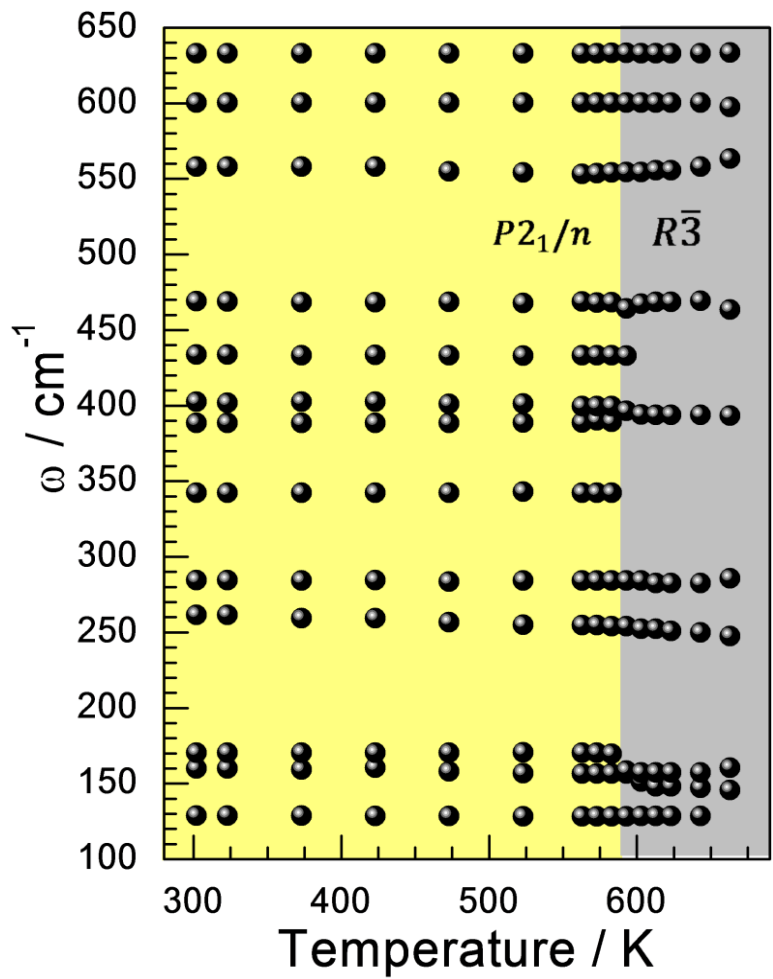

**Fig. 3** – Temperature dependence of the depicted TO phonon frequencies for all observed IR-active modes.

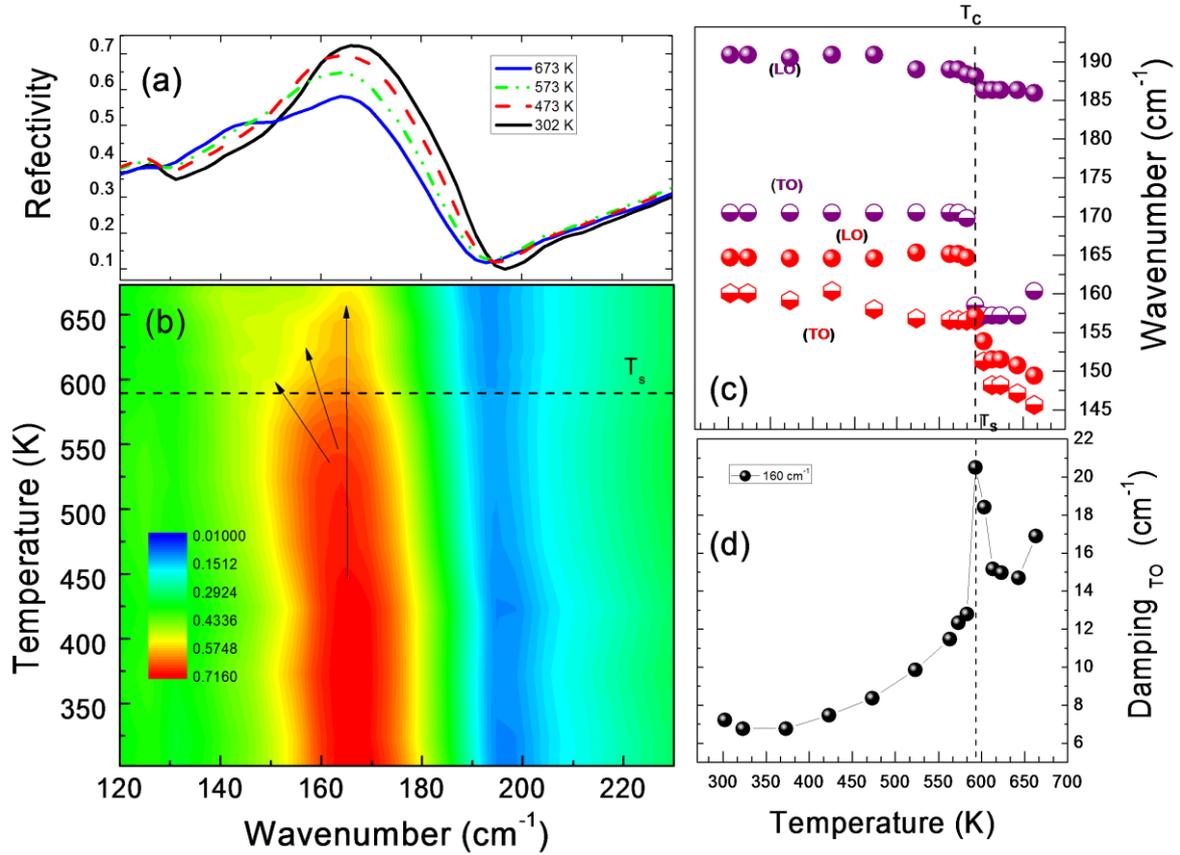

**Fig. 4** – (a) The reflection spectra, in the spectral region between 120 cm$^{-1}$ and 230 cm$^{-1}$, at several temperatures, and (b) the corresponding reflection intensity maps in the frequency-temperature axes for LCMO. (c) The temperature dependences of the TO and LO frequencies (from the modes #2 and #3) (d) and of the damping constants of the mode with ω$_{TO}$ ~ 160 cm$^{-1}$ (300 K).

Similarly, we have analyzed in detail the spectral region which includes the mode with $\omega_{TO}$~261 cm$^{-1}$ (mode #4), the second highest dielectric strength mode, as shown in Figure 5. In general, this mode has undergone a strong softening since the beginning of the heating of the sample, of about 8 cm$^{-1}$, before the structural transition, and more sharply, ca. 6 cm$^{-1}$, after the transition. Above $T_C$, the $\omega_{LO}$ component showed a strong softening of ~12 cm$^{-1}$, according to Figure 5(c). The deviation of this mode to low frequencies is well evidenced in the reflection spectra and in the corresponding reflection intensity maps in a spectral region between 240 cm$^{-1}$ and 330 cm$^{-1}$ [Figures 5(a) and 5(b)]. Also, the damping constant parameter for this mode presents a sharp peak in the vicinity of the critical point [Figure 5(d)]. It is worthy noticing that, at higher wavenumbers, the mode #11 ($\omega_{TO} \sim 558$ cm$^{-1}$) exhibits a strong anomalous hardening after the SPT, deviating about 11 cm$^{-1}$ from the room temperature value (Figure 3).

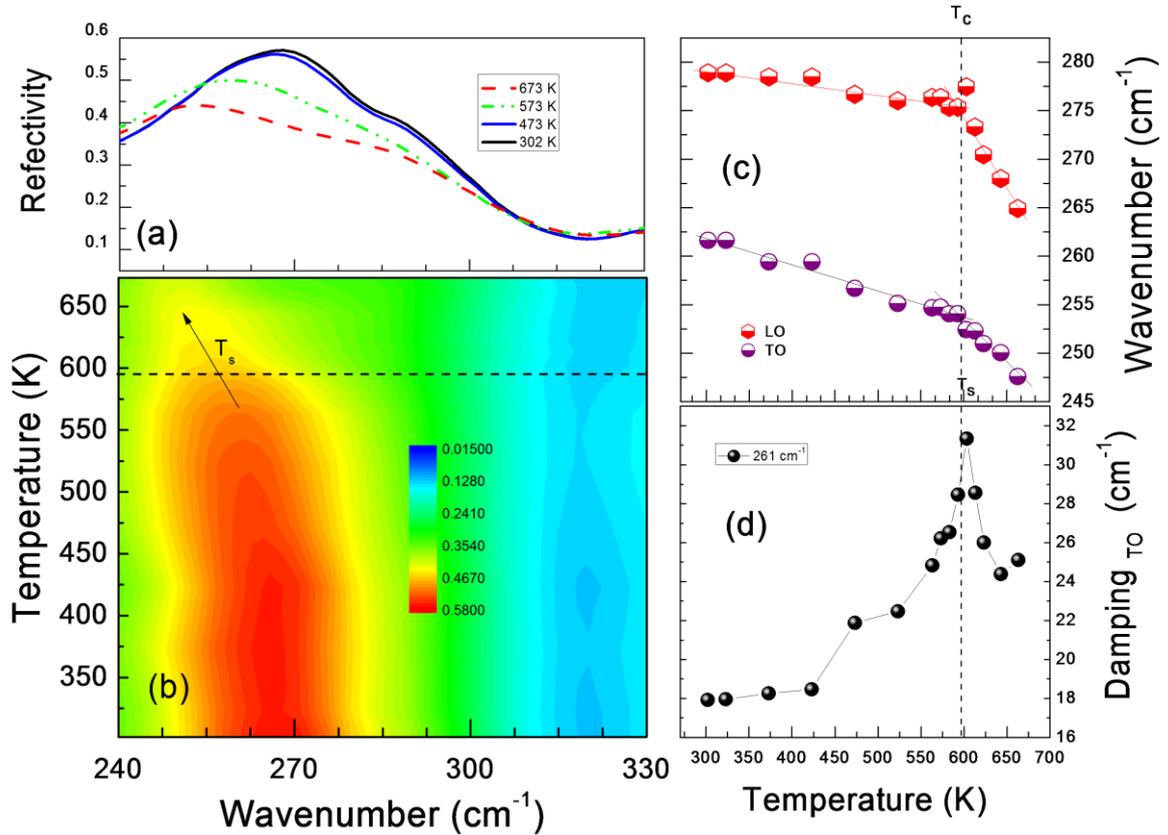

**Figure 5** – (a) The reflection spectra in the range between 240 cm$^{-1}$ and 330 cm$^{-1}$ at several temperatures and (b) the corresponding reflection intensity maps in the frequency-temperature axes for LCMO. (c) The temperature dependences of the TO and LO frequencies (d) and of the damping constants of the mode with $\omega_{TO} \sim 261$ cm$^{-1}$ (300 K).

Generally, the mode frequency shifts $\Delta\omega_{0i}$ may be caused by the change of crystal volume $\Delta V$ in the vicinity of the structural phase transition. Thus, it is expected that frequencies could be renormalized due to lattice thermal expansion and atom rearrangement effects, which cause a frequency shift of the form[40]

$$\delta\omega_j(T) = -\omega_j \int_0^T g_j(T)\alpha_V(T)dT \qquad (6)$$

where $g_j$ is the mode Grüneisen parameter (MGP) of the $j^{th}$ mode and $\alpha_V$ is the volumetric thermal expansion coefficient. In a first approximation, if the MGP is temperature independent, Eq. (6) may be rewritten as

$$\delta\omega_j(T) = -\omega_j g_j \frac{\Delta V(T)}{V} \qquad (7)$$

For LCMO there are no detailed data on thermal lattice expansion. However, Bull *et al.* [15] measured the lattice parameters at room temperature and at 673 K for LCMO. They obtained the cell volumes V(293 K) = 235.8416 Å$^3$ and V(673 K) = 237.5686 Å$^3$, respectively, which yields the ratio ΔV/V = 7.3x10$^{-3}$. Using this value and our experimental frequency data in the temperature range between 293 and 673 K, and supposing temperature-independent Grüneisen parameters, we can apply Eq. (7) to obtain a rough estimate of the MGP for the strongest dielectric modes (see Table IV). We observe that the obtained values are significantly higher in magnitude than those for quartz[40] or silicon clathrates, which are in the range $0.13 \lesssim g_j \lesssim 1.46$[41]. We also note that $g_j$ for the observed modes at around 160, 170 and 261 cm$^{-1}$ are the ones that most contributed to the thermal expansion in LCMO.

Table IV - Grüneisen parameters for some isolated vibrational modes of La$_2$CoMnO$_6$.

| Mode | $\omega_0$ (300 K) | $\omega_0$ (673 K) | $\frac{\Delta\omega_0}{\omega_0}$ | $\frac{\Delta V}{V}$ | $g$ |
|---|---|---|---|---|---|
| #2 | 160 | 145 | -0.093 | 0.0073 | 12.90 |
| #3 | 170 | 160 | -0.061 | 0.0073 | 8.33 |
| #4 | 261 | 247 | -0.055 | 0.0073 | 7.54 |
| #8  | 402 | 393 | -0.023 | 0.0073 | 3.19  |
| #11 | 558 | 563 |  0.009 | 0.0073 | -1.30 |

The high values obtained for the Grüneisen parameters indicate that, due to the structural transition, a deviation occurs from the quasi-harmonic approximation[42]. In this context, we observed a critical increase of the damping constants $\gamma(T)$ around the structural phase transition [see panels of Figs. 4d and 5d]. Such anomalous increase has been explained by anharmonic coupling of near modes, in which the linewidth diverge at the approach of the transition[43]. The anharmonic couplings between phonon modes create a complex self-energy shift:

$$\hbar\left[(\Delta\omega_j(\omega) + i\Gamma_j(\omega)\right], \tag{8}$$

in which the real part is the observed frequency shift, while the imaginary part is related to the phonon damping, $\gamma_j$, by $\Gamma_j(\omega) = \gamma_j \frac{\omega}{2\omega_j}$ [44]. This is expressed in a mutual repulsion of the modes frequencies observed at 160 and 170 cm$^{-1}$ (see Fig. 4c) and in an exchange of their dielectric strengths (see Figure 6), i.e., while the strength of the strongest mode #2 (~160 cm$^{-1}$) increases on heating up to $T_C$, and then decreases above this temperature, the dielectric strength of mode #3 (~170 cm$^{-1}$) has an opposite behaviour (becoming the strongest one at high temperatures). Similar behaviour has been observed for TbFe$_3$(BO$_3$)$_4$ and GdFe$_3$(BO$_3$)$_4$[43]. In addition, we observed an abrupt separation between the TO (170 cm$^{-1}$) and LO (191 cm$^{-1}$) branches frequencies of the mode #2 at $T_C$ [see Figure 4 (c)], which is associated with the broadening of the reflectivity band defined by them. A similar anomalous increase in the linewidth of the mode #4 (with ω$_{TO}$ ~ 261 cm$^{-1}$) was observed at the same temperature. Additionally, from Figure 6, we can see that the intrinsic dielectric constant remains always relatively low, even around the SPT. However, a subtle change of trend can be observed after the structural transition, i.e., the dielectric constant increases slightly from room temperature up to $T_C$; then, it starts decreasing above this temperature.

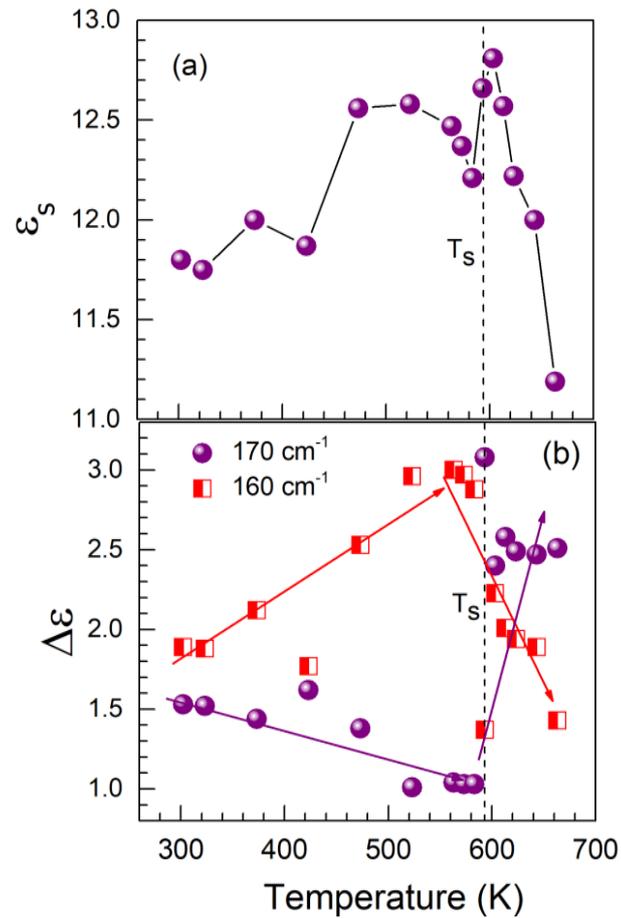

**Figure 6** – The temperature dependences of (a) the intrinsic dielectric constant, $\varepsilon_s$, and of (b) the oscillator strengths, $\Delta\varepsilon$, of the modes #2 (at 160 cm$^{-1}$) and #3 (at 170 cm$^{-1}$).

## 4. Conclusions

In this paper, we have investigated the temperature-dependence of the IR-active lattice phonon modes for a partially ordered magnetodielectric La$_2$CoMnO$_6$, in a temperature range from room temperature up to 673 K. Our results show a first-order structural phase-transition observed at around 590 K from the monoclinic structure, with $P2_1/n$ symmetry, to a rhombohedral phase, with $R\bar{3}$ symmetry. In opposition to recent reports, the partial ordering of the studied sample did not inhibit the structural phase transition. The modes with the highest contributions to the dielectric constant presented anomalous trends for the temperature dependences of the phonon parameters (TO and LO branches frequencies, oscillator strengths and damping constants) and they play an important role on the SPT. Also, the damping deviation at the vicinity of the critical temperature revealed a relevant lattice anharmonicity, especially for the low frequency modes.

**Acknowledgments**

The authors are grateful to CNPq, CAPES, Fapemig and Fapema for co-funding this work.